\documentclass[aps,prx,onecolumn,footinbib,11pt,notitlepage]{revtex4-1}
\usepackage{wrapfig}
\usepackage{amsmath}
\usepackage{graphicx}
\usepackage{dcolumn}
\usepackage{bm}
\usepackage{braket}
\usepackage{color}
\usepackage{enumitem}
\usepackage{hyperref}
\hypersetup{
	colorlinks=true,
	linkcolor=blue,
	filecolor=blue,
	citecolor=blue,  
	urlcolor=black,
}

\def\be{\begin{equation}}
	\def\ee{\end{equation}}
\def\ba{\begin{eqnarray}}
	\def\ea{\end{eqnarray}}

\usepackage{amsfonts}
\usepackage{subfigure}
\usepackage{bm}
\usepackage{xr}

\begin{document}
\title{
Noisy entanglement-assisted classical capacity as a security framework for two-way quantum key distribution protocols
}

\author{Quntao Zhuang$^{1,2}$}
\email{quntao@mit.edu}
\author{Zheshen Zhang$^1$}
\author{Jeffrey H. Shapiro$^1$}
\affiliation{
\footnotesize
$^1$Research Laboratory of Electronics, 
Massachusetts Institute of Technology, Cambridge, Massachusetts 02139, USA\\
$^2$Department of Physics, 
Massachusetts Institute of Technology, Cambridge, Massachusetts 02139, USA}
\maketitle
Quantum key distribution (QKD) offers unconditional security against eavesdropping~\cite{Bennett20147}, but state-of-the-art secret key rates (SKRs) for QKD are only $\sim$1\,Mbps on a 50-km-long fiber link~\cite{lucamarini2013efficient}, i.e., orders of magnitude lower than classical fiber-communication rates. Floodlight QKD (FL-QKD) is a Gaussian two-way quantum key distribution protocol (TW-QKD) that is theorized to be capable of Gbps SKRs over metropolitan-area distances~\cite{Quntao_2015}.  In FL-QKD, Alice sends quantum states to Bob.   While they estimate the correlation between eavesdropper Eve and Bob, Bob encodes a raw key on the light he receives and sends that modulated light back to Alice.  Unfortunately, until now FL-QKD's security proof is limited to frequency-domain collective attacks~\cite{Quntao_2015}.  More generally, the security of TW-QKD against coherent attacks is still an open problem. 

In this paper, we use the noisy entanglement-assisted capacity~\cite{Zhuang_2016_cl} to create a coherent-attack security framework for Gaussian TW-QKD protocols in the asymptotic region.  We use Eve's disturbance of Alice and Bob's Gaussian-state covariance matrix---which can be bounded from homodyne measurements---to quantify her intrusion on a Gaussian TW-QKD protocol, such as those in Refs.~\cite{Pirandola_2008,Quntao_2015}, and obtain therefrom unconditional security against a coherent attack. Our results pave the way towards high-rate QKD with unconditional security.
\begin{wrapfigure}[11]{l}{0.35\textwidth}
\includegraphics[width=0.35\textwidth]{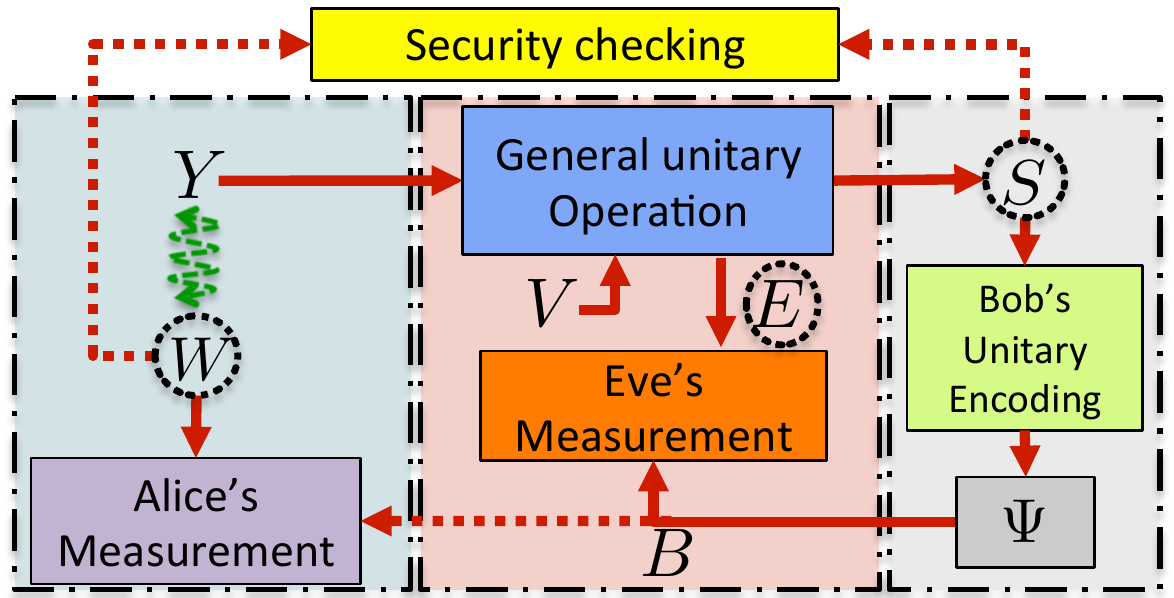}
\caption[caption]{Single-use schematic of a Gaussian TW-QKD protocol.}
\label{scheme_QKD}
\end{wrapfigure}
\ \ \ \ \ {\em Gaussian TW-QKD protocols.---}
Figure.~\ref{scheme_QKD} shows a single use of a general Gaussian TW-QKD protocol.  Alice prepares a signal-reference pair $(Y,W)$ in a two-mode squeezed vacuum (TMSV) state with mean photon number $N_S$.  She measures a portion of $W$ for security checking, and sends the signal $Y$ to Bob through a forward channel that is controlled by Eve.  In general, Eve performs a unitary operation on $Y$ and her pure-state input $V$, retaining its $E$ output and delivering its signal output $S$ to Bob.  (Note that $V$ and $E$ can have multiple modes per channel use.)  In Eve's coherent attack, her unitary operation can act jointly on \emph{all} channel uses~\cite{RMP_security}. 

\begin{wrapfigure}[10]{l}{0.35\textwidth}
\includegraphics[width=0.35\textwidth]{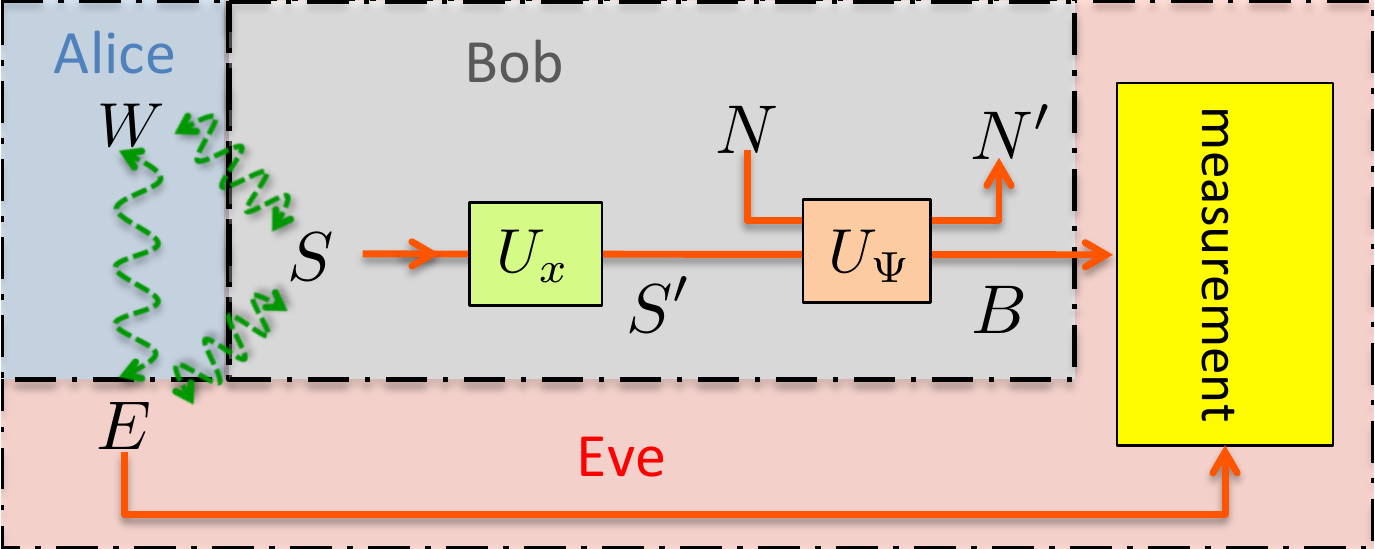}
\caption[caption]{Gaussian TW-QKD protocol from Eve's perspective.}
\label{scheme_channel}
\end{wrapfigure}

Figure~\ref{scheme_channel} contains a schematic plot of the protocol after Bob receives $S$.    
He measures a portion of $S$ for security checking, and encodes a random symbol $x$ on the remainder.  Alice and Bob's security checking uses homodyne measurements~\cite{Cov1,Cov2} to estimate constraints on the covariance matrix of the joint state $\hat{\rho}_{SW}$; in the asymptotic regime these estimates will be perfect.  Bob encodes $x$ with a unitary $\hat{U}_x$ composed of a phase shift $\theta_x$ and a displacement $d_x$ that are easily realized with linear optics.
Conditioned on the message $x$, the encoded mode has annihilation operator
$
\hat{a}_S^{\prime\left(x\right)}=e^{i\theta_x}\hat{a}_S+d_x.
$
The $d_x$'s are assumed to be zero-mean Gaussian random variables, implying that encoding on a vacuum state will average to produce a thermal state.
The encoding scheme is symmetric, i.e., 
$
\sum_x P_X\left(x\right) e^{i\theta_x}d_x=0
$, and energy constrained, viz., $E_X=\sum_x P_X\left(x\right) |d_x|^2$.
Thus it includes the random-displacement encoding scheme used in Refs.~\cite{Pirandola_2008,Ottaviani_2015} and the phase encoding employed in FL-QKD~\cite{Quntao_2015,zhuang2017large}. The unconditional state of $\left(S^\prime,W\right)$ is non-Gaussian in general.

Bob's encoded signal passes through channel $\Psi$ that models the part of the return channel that is not under Eve's control, e.g., loss in Bob's terminal, but we will allow $\Psi$ to be any Gaussian channel without excess noise:  a pure-loss channel (transmissivity $\eta$), a quantum-limited amplifier (gain $G_B$), or a quantum-limited phase conjugator (gain $G_B$).  After $\Psi$, Bob sends its output $B$ to Alice through a channel controlled by Eve.   Alice jointly measures the light she receives with part of $W$ to obtain a raw key from which the secret key will be distilled after Alice and Bob use their covariance-matrix constraints to bound the information gained by Eve.

{\em Bounding eavesdropper's information gain.---}
In the asymptotic regime, a QKD protocol's secret-key efficiency (SKE), in bits per channel use, against a coherent attack is given by the Devetak-Winter formula~\cite{devetak2005distillation,RMP_security} ${\rm SKE}= \max\left[\xi I_{AB}-I_E,0\right]$,
where $I_{AB}$ is Alice and Bob's 
Shannon information in bits per channel use, $\xi$ is their reconciliation efficiency, and $I_E$ is Eve's Holevo-information gain in bits per channel use.  (Note that Alice and Bob's SKR equals $R$\,SKE, where $R$ is Bob's symbol rate.)  The maximization in the Devetak-Winter formula needs to be performed over all possible attacks that pass the security checking measurements, i.e., that are consistent with Alice and Bob's measured covariance-matrix constraints.  We will perform that maximization on $\chi_E\equiv I_E/M_E$, Eve's Holevo information in bits per mode, where $M_E$ is the number of modes used per encoded symbol. Thus, because $\xi I_{AB}$ can be inferred from Alice and Bob's reconciliation step, the asymptotic security proof of the TW-QKD protocols rests on putting an upper bound on $\chi_E$.

Bounding $\chi_E$ for a TW-QKD protocol is complicated by Eve's simultaneously attacking the forward and backward channels~\cite{Pirandola_2008,two_way_no_loss,Han_2014,generalization_no_loss,Quntao_2015,Ottaviani_2015}.
Consequently, the usual techniques, such as the entropic uncertainty principle~\cite{berta2010uncertainty}, are not applicable here because of loss. 
Recognizing that the TW-QKD protocol shown in Fig.~\ref{scheme_channel} can be regarded as noisy entanglement-assisted classical communication from Bob to Eve, we use the noisy entanglement-assisted classical capacity formula~\cite{Zhuang_2016_cl} to place on upper bound on $\chi_E$.  Thus we establish a new security framework for TW-QKD protocols.
Consider a multiple channel uses QKD session over $M$ mode pairs. We use the same notation as Fig.~\ref{scheme_QKD} with subscripts indicating the different mode pairs, i.e., ${\bf S}=S_1S_2\cdots S_M$, ${\bf W}=W_1W_2\cdots W_M$, and ${\bf B}=B_1B_2\cdots B_M$. For Gaussian protocols, the $\hat{U}_x$'s are
\begin{wrapfigure}[13]{l}{0.53\textwidth}
\centering
\subfigure[\,TMSV protocol with random displacement.]{
\includegraphics[width=0.24\textwidth]{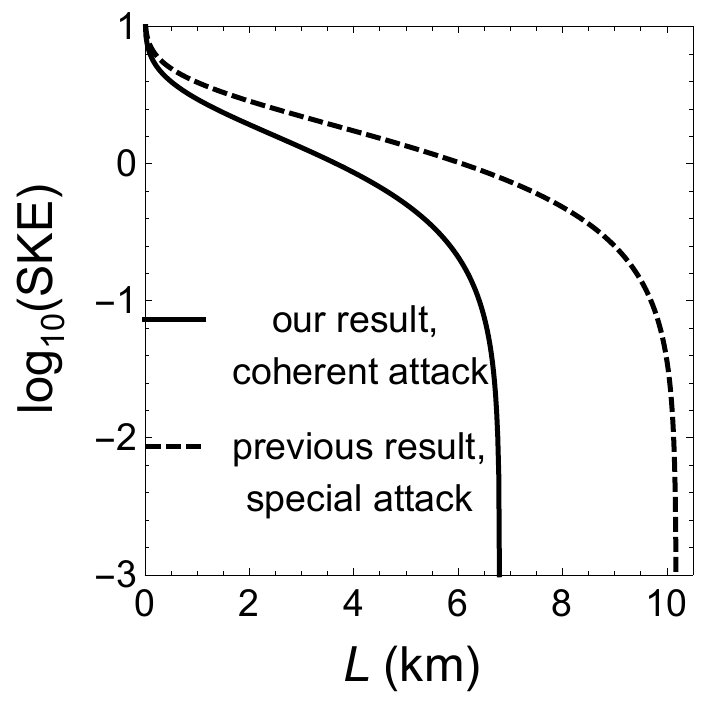}
\label{rate1}
}
\centering
\subfigure[\,FL-QKD protocol. $G_B=10^6$, $N_S$ is optimized.]{
\includegraphics[width=0.24\textwidth]{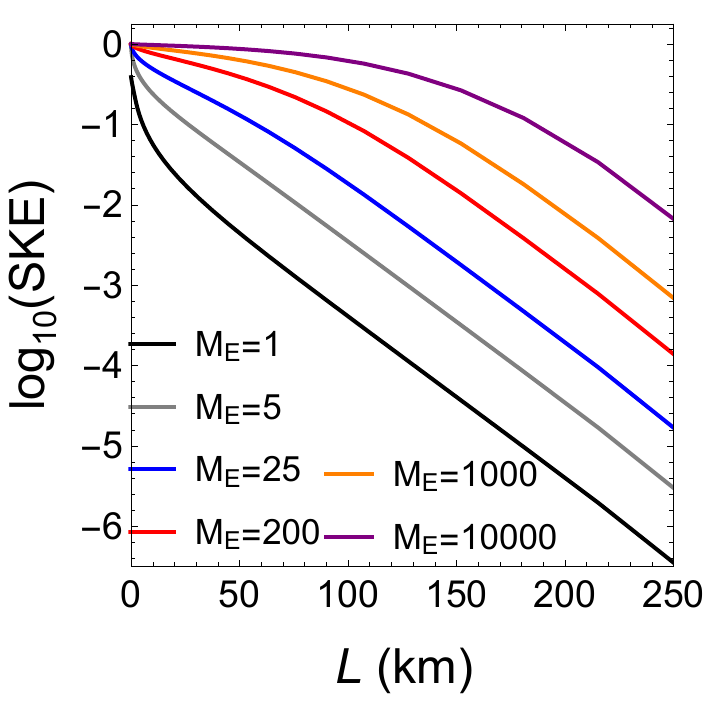}
\label{rate2}
}\vspace*{-.1in}
\caption[caption]{Secret-key rates versus path length $L$.}
\label{rate}
\end{wrapfigure} 
covariant with $\Psi$, thus Eve's information gain is upper bounded by a maximization, given the covariance-matrix constraints, over multiple mode pairs, i.e., the multi-letter formula~\cite{Zhuang_2016_cl},
\begin{align}
\chi_E^{\left(M\right)} &= \max_{\hat{\rho}_{\bf SW}}
F\left[\hat{\rho}_{\bf SW}\right], \mbox{ with }
\label{capacity_Eve}
\\
F\left[\hat{\rho}_{\bf SW}\right]&\equiv
S\left(\bf \hat{\rho}_B\right)-E_{{\left(\Psi^{\otimes M}\right)}^c\otimes \mathcal{I}}\left[\hat{\rho}_{\bf SW}\right],
\label{capacity_covariant}
\end{align}
where each $\hat{\rho}_{B_m}=\sum_x P_X\left(x\right) \Psi[\hat{U}_x^\dagger \hat{\rho}_{S_m}\hat{U}_x^\dagger]$, i.e., we have assumed independent encoding on each mode pair. With dependent encoding, Eq.~(\ref{capacity_Eve}) is still an upper bound. A trace-preserving completely-positive map $\phi$ has complementary channel we denote as $\phi^c$, and the entropy gain of $\phi$ on state $\hat{\rho}$ is $
E_\phi \left[\hat{\rho}\right]\equiv S\left(\phi\left[\hat{\rho}\right]\right)-S\left(\hat{\rho}\right)
$. To reduce Eq.~(\ref{capacity_Eve}) to a single mode-pair (single-letter) formula, we use the subadditivity of $F\left[\hat{\rho}_{\bf S W}\right]$~\cite{Zhuang_2016_cl},  and, because $\Psi$ is a Gaussian channel, this also ensures that the maximum of Eq.~(\ref{capacity_Eve}) is achieved by a Gaussian-state $\hat{\rho}_{\bf SW}$~\cite{Zhuang_2016_cl} under the given covariance-matrix constraints.

At this point we introduce the covariance-matrix constraints that Alice and Bob will obtain from their security checking.  The first will be the total mean photon number of the signal received by Bob, $
\sum_{n=1}^M
\braket{\hat{a}^\dagger_{S_n}\hat{a}_{S_n}}=M\overline{\kappa}_S N_S
$. The second will be the total cross correlation between Alice's retained and Bob's received modes, 
$
{\sum_{m,n=1}^M( |\braket{\hat{a}_{S_m}\hat{a}_{W_n}}|^2+|\braket{\hat{a}_{S_m}\hat{a}_{W_n}^\dagger }|^2)}=\left(1-f_E\right)\overline{\kappa}_S {MN_S\left(N_S+1\right)}
$. Here, $f_E, \overline{\kappa}_S$ quantify Eve's intrusion on the quantum channels, and $0\le f_E\le 1$, required by physics.  By constraining the covariance matrix we can bound $\chi_E$, because $\chi_E$ decreases with increasing $\overline{\kappa}_S$ and it increases with increasing $f_E$. The total mean photon number constrains the covariance matrix's diagonal elements, while $\sum_{m,n=1}^M (|\braket{\hat{a}_{S_m}\hat{a}_{W_n}}|^2+|\braket{\hat{a}_{S_m}\hat{a}_{W_n}^\dagger }|^2)\ge |\sum_{n=1}^M \braket{\hat{a}_{S_n}\hat{a}_{W_n}}|^2+|\sum_{n=1}^M \braket{\hat{a}_{S_n}\hat{a}_{W_n}^\dagger }|^2
$ implies that the covariance matrix's off-diagonal elements give a lower bound on the total correlation.

By using optimization techniques similar to those in Ref.~\cite{Quntao_2015}, we can show that our constraints permit Eq.~(\ref{capacity_Eve}) to be reduced to a single-letter formula that can be evaluated as a function of the intrusion parameters $\overline{\kappa}_S, f_E$. With Eve's information gain in hand, the SKE can then be obtained from the Devetak-Winter formula. In the examples that follow, we will use $\overline{\kappa}_S$ equal to the one-way fiber loss $\kappa_S=10^{-0.02 L}$ that Alice and Bob will see when they are connected by $L$\,km of fiber.

{\em TMSV protocol with random displacement{\rm~\cite{Pirandola_2008,Ottaviani_2015}}.---} In this protocol, Alice has access to the full TMSV, Bob encodes each mode using random displacements with power $E_X$, and $\Psi$ is the noiseless identity channel.  Figure~\ref{rate1} compares our SKE lower bound with the SKE result from Refs.~\cite{Pirandola_2008,Ottaviani_2015} when $f_E=0, \xi=1$ and  $E_X\gg1, N_S\gg1$.  Our lower bound, which applies for a coherent attack in the asymptotic regime, is much lower than the one from~Refs.~\cite{Pirandola_2008,Ottaviani_2015}, which only applies for a special class of collective attacks.  We believe that much of this gap is due to our giving Eve all the light on the backward channel, which is an overly conservative assumption given the short distances involved, e.g., $\kappa_S=0.63$ for $L=10$\,km.
For TW-QKD protocols like FL-QKD, which are capable of long-distance operation, we expect that our SKE lower bound will be
tighter at those long distances, e.g., when $\kappa_S = 0.1$ for $L=50$\,km.

{\em FL-QKD protocol{\rm ~\cite{Quntao_2015,zhang2016floodlight,zhuang2017large}}.---}
FL-QKD offers Gbps SKRs at long distances by virtue of three features.  First, Alice uses low-brightness amplified spontaneous emission light (ASE), together with TMSV light, in her transmission to Bob, while retaining a high-brightness ASE reference as a homodyne-detection local oscillator for measuring Bob's encoded message.  Nevertheless, even with only partial access to 
 the purification $W$, Alice can still establish asymptotic security. 
Second, Bob uses a high-gain ($G_B \gg 1$) amplifier as his $\Psi$, which overcomes the backward-channel loss issue that plagues previous TW-QKD protocols~\cite{Pirandola_2008,Ottaviani_2015}.
Finally, Bob uses multi-mode encoding, $M_E \gg 1$, that allows Alice to decode Bob's message despite the low-brightness of the signal light she transmitted.  Previous work~\cite{Quntao_2015,zhang2016floodlight,zhuang2017large} has only proven Fl-QKD's security against a  frequency-domain collective attack. Here we apply our framework to obtain its asymptotic SKE against a coherent attack.

FL-QKD, uses phase encoding, so its $E_X=0$.  Although alphabets larger than binary are known to be beneficial~\cite{zhuang2017large}, here we will consider binary encoding with phases $\theta_0=0, \theta_1=\pi$ representing the bit values 0 and 1. Figure~\ref{rate2} plots FL-QKD's SKE against a coherent attack in the asymptotic regime assuming $f_E=0, \xi=1$ for a variety of $M_E$ values where we have optimized over the source brightness at each distance. 
The red line corresponds to the operating point of $M_E=200$ as used in Refs.~\cite{Quntao_2015,zhuang2017large} for the frequency-domain collective attack. We see that with $M_E\gg1$ and $R=10$\,Gbps, FL-QKD provides Gbps SKRs at long distances. Note that FL-QKD's SKE against the coherent attack---as determined here---coincides with the SKE obtained in Ref.~\cite{Quntao_2015} against the frequency-domain collective attack, and hence the SKE incurred with $f_E>0$ for a coherent attack can be found from that reference.

%

\end{document}